\documentclass[pra, twocolumn, eqsecnum]{revtex4}
\usepackage{epsfig,amsmath,amssymb,bm,epsf,graphicx,latexsym,amsfonts,times}

\newcommand {\psig}{\Sigma_}
\newcommand {\pxi}{\Xi_}
\newcommand {\apsig}[1]{\langle \Sigma_{#1} \rangle}
\newcommand {\apxi}[1]{\langle \Xi_{#1} \rangle}
\newcommand {\apsigxi}[2]{\langle \Sigma_{#1} \Xi_{#2} \rangle}

\begin{document}

\title{One qubit almost completely reveals the dynamics of two}

\author{Thomas F. Jordan}
\email[email: ]{tjordan@d.umn.edu}
\affiliation{Physics Department, University of Minnesota, Duluth, Minnesota
55812} 
\author{Anil Shaji}
\email[email: ]{shaji@unm.edu}
\affiliation{The University of New Mexico, Department of Physics and
Astronomy, 800 Yale Blvd. NE, Albuquerque, New Mexico 87131} 
\author{E. C. G. Sudarshan}
\email[email: ]{sudarshan@physics.utexas.edu}
\affiliation{The University of Texas at Austin, Center for Statistical
Mechanics, 1 University Station C1609, Austin Texas 78712}   

\begin{abstract}
From the time dependence of states of one of them, the dynamics of two
interacting qubits is determined to be one of two possibilities that differ
only by a change of signs of parameters in the Hamiltonian. The only
exception is a simple particular case where several parameters in
the Hamiltonian are zero and one of the remaining nonzero parameters has no
effect on the time dependence of states of the one qubit. The mean values
that describe the initial state of the other qubit and of the correlations
between the two qubits also are generally determined to within a change of
signs by the time dependence of states of the one qubit, but with many more
exceptions. An example demonstrates all the results. Feedback in the equations
of motion that allows time dependence in a subsystem to determine the dynamics
of the larger system can occur in both classical and quantum mechanics. The role
of quantum mechanics here is just to identify qubits as the simplest objects to
consider and specify the form that equations of motion for two interacting
qubits can take. 
\end{abstract}



\maketitle

\section{Introduction \label{sec1}}

What can we learn about the dynamics of two interacting qubits by observing
the time dependence of states of one of them? How much can the dynamics of
the two be changed without changing the time dependence of states of the
one? We will show that the dynamics of the two can be changed only by a
change of signs of parameters in the Hamiltonian. There are only two simply
related possibilities for the dynamics of the two qubits that give the same
time dependence for the states of the one. This is set out in sections
\ref{sec2}-\ref{sec4}. 

There is only one exception: in a simple particular case where several
parameters in the Hamiltonian are zero, one of the remaining nonzero
parameters can vary over the whole range of real numbers without changing
the time dependence of states of the observed qubit. This is described in
Section~\ref{sec5}. 

Determining the dynamics of the two qubits generally takes three time
derivatives at time zero, expansion to third order in powers of time, of
the mean values that describe the states of the observed qubit evolving in
time, but it can take six time derivatives when some of the parameters in
the Hamiltonian are zero.

The mean values that describe the initial state of the unobserved qubit and
of the correlations between the two qubits also are generally determined to
within a change of signs by the time dependence of states of the observed
qubit. This generally takes four time derivatives. There are many
exceptions. This is described in Section \ref{sec6}. An example that
demonstrates all the results is worked out in Section \ref{sec67}.

Various implications and applications can be considered. One broad view of our
results, framed only by the context of open quantum dynamics of a system
evolving together with its environment, is that time dependence in the system
can provide almost complete information about the environment and the
interaction of the system with the environment, without measurements of the
environment \cite{devitt06a,cole06a}. Our results provide an example that
demonstrates the general statement. They have immediate application when the
system is a qubit and its interaction with its environment can be modeled by
interactions with qubits \cite{ecgs03b}.

A more specific application can be seen in quantum information processing. When
a physical device's performance of an operation is tested to verify that an
interaction between two qubits is what it was designed to be, complete quantum
process tomography
\cite{chuang97a,poyatos97a,nielsen00a,childs01a,leung03a,altepeter03a,
myrskog03a,obrien04a,weinstein04a,howard05a} is not needed. Our results suggest
a simpler procedure. The dynamics of the two qubits can be determined almost
completely by measurements of time dependence for one qubit with varying
initialization of the state for that qubit but not for the two qubits.  

The time dependence of the states of the one qubit is an example of an open
quantum dynamics described by maps of states of a subsystem caused by
unitary Hamiltonian dynamics in a larger system. Since these maps generally
do not depend on time as a semigroup, it is an open question how and to
what extent a map at one time or at several times determines the map at
other times. Here we give an answer for the dynamics of two qubits. The
maps of the states of the one qubit for all times are determined by the
Hamiltonian for the two qubits and the initial correlations between the two
qubits \cite{jordan04a,jordan06a}. We show how and to what extent these are
determined by the maps in a neighborhood of the initial time.

A classical analog, considered in Section \ref{sec7}, exhibits the logical
structure of our method in a setting where it can be easily seen. Feedback in
the equations of motion is what lets time dependence in a subsystem determine
the dynamics of the larger system. This can happen in both classical and quantum
mechanics. Here, quantum mechanics simply identifies qubits as the simplest
objects to consider and specifies the form that equations of motion for two
interacting qubits can take.  

\section{Method \label{sec2}}

We consider two qubits, one described with Pauli matrices $\psig1$,
$\psig2$, $\psig3$ and the other with Pauli matrices $\pxi1$, $\pxi2$,
$\pxi3$. We assume the two qubits interact with each other but not with
anything else. The time dependence is generated by a Hamiltonian 
\begin{equation}
H= \frac{1}{2} \alpha_j \psig{j} + \frac{1}{2}\beta_k \pxi{k} + \frac{1}{2}
\gamma_{jk} \psig{j}\pxi{k} \label{eq:hamilfirst}
\end{equation}
with real parameters $\alpha_j$, $\beta_k$, $\gamma_{jk}$.
When an index $j$ or $k$ is repeated, there is to be a sum over the values
$1$, $2$, $3$ for that index. 

In almost all cases, the dynamics of the two qubits is almost completely
determined by the time dependence of states of the $\Sigma $ qubit alone.
To see this, we look at the mean values 
\begin{eqnarray}
\apsig{n}(t) &=& \big\langle e^{itH} \psig{n} e^{-itH} \big\rangle \nonumber \\
& =& u_{nj}(t)\apsig{j} + v_{nk}(t) \apxi{k} + w_{njk} \apsigxi{j}{k} \qquad
\;\;
\label{eq:mean}
\end{eqnarray}
for $n=1, \, 2, \, 3$ which describe the states of the $\Sigma $ qubit at
time $t$. We consider variable initial states of the $\Sigma $ qubit
described by variable mean values $\apsig{j}$. The $u_{nj}(t)$ are
determined by the $\apsig{n}(t)$ for variable $\apsig{j}$. We will see that
the dynamics for the two qubits is almost completely determined by the
$u_{nj}(t)$. Later we will see that generally the $\apxi{k}$ and
$\apsigxi{j}{k}$ also are almost completely determined, so the initial
state of the $\Xi $ qubit and of the correlations between the two qubits
is almost completely determined by the time dependence of states of the
$\Sigma $ qubit, but nothing more can be learned about the dynamics from
the $\apxi{k}$, $\apsigxi{j}{k}$ and $v_{nk}(t)$, $w_{njk}(t)$.

The same dynamics may be described by different $\psig{j}$, $\pxi{k}$.
There may be changes of the $\psig{j}$, $\pxi{k}$ that do not change the
Hamiltonian, or that do change the way the Hamiltonian is expressed in
terms of the $\psig{j}$, $\pxi{k}$ but do not change the dynamics. We will
not be concerned with the differences made by these changes.

The time dependence of states of the $\Sigma $ qubit can be described
\cite{jordan04a,jordan06a} by maps of the mean values $\apsig{j}$. Of the
$\apsig{j}$, $\apxi{k}$, $\apsigxi{j}{k}$, the $\apsig{j}$ describe the
state of the $\Sigma $ qubit, and the $\apxi{k}$ and $\apsigxi{j}{k}$ are
considered to be parameters of the maps that describe how the dynamics of
the two qubits drives the evolution of the $\Sigma $ qubit
\cite{jordan04a,jordan06a}. Different $\apxi{k}$ or $\apsigxi{j}{k}$
specify different maps. Each map applies to variable states of the
$\Sigma $ qubit described by  variable $\apsig{j}$. In almost all cases,
the dynamics of the two qubits is almost completely determined by any one
of these maps.

\section{Magnitudes \label{sec3}}

We calculate $u_{nj}(t)$ as a series in powers of $t$ by identifying the
coefficients of $\psig{j}$ in the power series for $e^{itH} \psig{n}
e^{-itH}$. In first order we get
\begin{equation}
-i \big[\psig1, \, H \big] = \alpha_2 \psig3 - \alpha_3 \psig2+
\gamma_{2k} \psig3 \Xi_k - \gamma_{3k} \psig2 \Xi_k 
\label{eq:evol1}
\end{equation}
for $n=1$. The coefficients of the $\psig{j}$ in this and the similar
equations for $-i \big[\psig2, \, H \big]$ and $-i\big[\psig3, \, H \big]$
determine $\alpha_1$, $\alpha_2$, $\alpha_3$. 

In second order, the $\psig{j}$ terms in $-i \big[ -i \big[ \psig1, \, H
\big], \, H \big]$ are
\begin{eqnarray}
-\big[ (\alpha_2)^2 + (\alpha_3)^2 + \gamma_{2k}\gamma_{2k} + \gamma_{3k}\gamma_{3k}
\big] \psig1 \hspace{1.5 cm} \nonumber \\
+ \big[ \alpha_1 \alpha_2 + \gamma_{1k}\gamma_{2k}\big] \psig2 + \big[ \alpha_1
\alpha_3 + \gamma_{1k}\gamma_{3k}\big] \psig3.
\label{eq:evol2}
\end{eqnarray}
These and the $\psig{j}$ terms in $-i \big[ -i \big[ \psig2, \, H \big],
\, H \big]$ and \newline $-i\big[-i\big[\psig3, \, H \big], \, H \big]$
determine
\begin{equation}
\label{dots}
\gamma_{1k}\gamma_{1k}, \; \gamma_{2k}\gamma_{2k}, \; \gamma_{3k}\gamma_{3k}, \;
 \gamma_{1k}\gamma_{2k}, \;  \gamma_{2k}\gamma_{3k}, \; \gamma_{3k}\gamma_{1k}
\end{equation}
which are the dot products of the vectors
\begin{eqnarray}
\label{vectors}
\vec \gamma_1 & = & (\gamma_{11}, \; \gamma_{12}, \; \gamma_{13}) \nonumber \\
\vec \gamma_2 & = & (\gamma_{21}, \; \gamma_{22}, \; \gamma_{23})\nonumber \\
\vec \gamma_3 & = & (\gamma_{31}, \; \gamma_{32}, \; \gamma_{33}).
\end{eqnarray}
Could anything more about the $\gamma_{jk}$ be determined? The Hamiltonian is
not changed if these three $\gamma $ vectors (\ref{vectors}) and the vector
$(\beta_1 ,\beta_2 ,\beta_3 )$ are all changed by the same rotation when the
Pauli matrices $\pxi1$, $\pxi2$, $\pxi3$ are changed the same way. The dynamics
is not changed for either qubit or for the system of two qubits by this change
of the $\pxi1$, $\pxi2$, $\pxi3$ used to describe the $\Xi$ qubit. In
particular, the mean values $\apsig{n}(t)$ are not changed.

The sign of $\vec \gamma_1 \cdot \vec \gamma_2 \times \vec \gamma_3 $ is not
determined, and it is not changed by rotations, so it is not changed when the
$\pxi1$, $\pxi2$, $\pxi3$ are changed. For the same $\apsig{n}(t)$, there could
be different possibilities for the dynamics of the two qubits corresponding to
the two different possible signs of $\vec \gamma_1 \cdot \vec \gamma_2 \times
\vec \gamma_3 $. We will consider these as two separate cases. We will find no
equations that connect them. The sign of $\vec \gamma_1 \cdot \vec \gamma_2
\times \vec \gamma_3 $ can keep the same value in all the equations, either
always plus or always minus. It is a free parameter. It is not determined by the
$\apsig{n}(t)$.

For each sign of $\vec \gamma_1 \cdot \vec \gamma_2 \times \vec \gamma_3 $, the
$\gamma_{jk}$ are determined to within rotations corresponding to changes of
$\pxi1$, $\pxi2$, $\pxi3$. We show in the Appendix that for each sign of $\vec \gamma_1 \cdot \vec \gamma_2 \times \vec \gamma_3 $, rotations of the $\psig{j}$ and $\pxi{k}$ can put the Hamiltonian in the form
\begin{equation}
H= \frac{1}{2} \alpha_j \psig{j} + \frac{1}{2}\beta_k \pxi{k} + \frac{1}{2}
\gamma_k \psig{k}\pxi{k} \label{eq:hamil}
\end{equation}
with real parameters $\alpha_j$, $\beta_k$, $\gamma_k$. This change uses a rotation of the $\psig{j}$ and $u_{nj}(t)$ that can be found
with knowledge of the dot products (\ref{dots}) that have been determined.
For any dynamics of the two qubits, described by a Hamiltonian
(\ref{eq:hamilfirst}), we can learn enough from the time dependence of states of
the $\Sigma $ qubit to change to a Hamiltonian of the form (\ref{eq:hamil}) and make the required change of of the $\psig{j}$ and $u_{nj}(t)$. From here on we will use this
simpler Hamiltonian form to describe the time dependence.
The changes made to put the Hamiltonian in the form (\ref{eq:hamil}) do not
change the value of $\vec \gamma_1 \cdot \vec \gamma_2 \times \vec \gamma_3 $, which gets called $\gamma_1  \gamma_2 \gamma_3 $.
This also is shown in the Appendix.

We will find that when $\vec \gamma_1 \cdot \vec \gamma_2 \times \vec \gamma_3
$ is zero, there still can be two possibilities for the dynamics of the two
qubits that give the same time dependence for states of the $\Sigma $ qubit.
They differ by a change of the signs of $\beta_2 $ and $\beta_3 $. When the sign
that distinguishes the two possibilities is specified, the dynamics of the two
qubits and the initial state of the $\Xi $ qubit and of the correlations
between the two qubits are almost always completely determined by the
$\apsig{n}(t)$. We have seen that $\alpha_1$, $\alpha_2$, $\alpha_3$ are
determined in first order and that $(\gamma_1)^2$, $(\gamma_2)^2$, 
$(\gamma_3)^2$ are determined in second order.

In third order, in $-i[ \;, \, H]$ applied to $\psig1$ three times, the
$\psig{j}$ terms that involve things not already determined are
\begin{equation}
\gamma_1\gamma_2 \beta_3 \psig2-\gamma_3\gamma_1\beta_2 \psig3.
\label{eq:evol3}
\end{equation}
These and the similar terms in $-i[ \;, \, H]$ applied to $\psig2$ and
$\psig3$ three times determine $\gamma_1\gamma_2\beta_3$, $\gamma_2\gamma_3
\beta_1$, $\gamma_3\gamma_1\beta_2$. We consider representative cases:
\begin{enumerate}
\item[$(i)$] None of $\gamma_1$, $\gamma_2$, $\gamma_3$ are zero;
\item[$(ii)$] $\gamma_1=0$, $\gamma_2 \neq 0$, $\gamma_3 \neq 0$;
\item[$(iii)$] $\gamma_1=0$, $\gamma_2=0$, $\gamma_3 \neq 0$.
\end{enumerate}
In case $(i)$, the magnitudes of $\beta_1$, $\beta_2$, $\beta_3$ are
determined. In case $(ii)$, the magnitude of $\beta_1$ is determined. In
case $(iii)$, nothing new is determined. We will consider signs in Section
\ref{sec4}.

In fourth order, in $-i[ \;, \, H]$ applied to $\psig1$ four times, the
$\psig{j}$ terms that involve things not already determined are
\begin{eqnarray}
(\gamma_3)^2 \big[ (\beta_1)^2 + (\beta_2)^2 \big] \psig1 +
(\gamma_2)^2 \big[ (\beta_3)^2 + (\beta_1)^2 \big] \psig1 \qquad \nonumber \\
+\gamma_1 \gamma_2 \beta_1 \beta_2 \psig2 + \gamma_1 \gamma_3 \beta_1
\beta_3 \psig3. \quad
\label{eq:evol4}
\end{eqnarray}
These and the similar terms in $-i[ \;, \, H]$ applied to $\psig2$ and
$\psig3$ four times determine 
\begin{eqnarray*}
(\gamma_3)^2 \big[ (\beta_1)^2 + (\beta_2)^2 \big], \quad
\gamma_1\gamma_2\beta_1 \beta_2, \\
(\gamma_2)^2 \big[ (\beta_3)^2 + (\beta_1)^2 \big], \quad \gamma_2 \gamma_3
\beta_2 \beta_3, \\
(\gamma_1)^2 \big[ (\beta_2)^2 + (\beta_3)^2 \big], \quad \gamma_3 \gamma_1
\beta_3 \beta_1.
\end{eqnarray*}
In case $(ii)$, all the magnitudes of $\beta_1$, $\beta_2$, $\beta_3$ are
now determined, In case $(i)$ they were already determined.

The rest of this section is for case $(iii)$. There only
$(\beta_1)^2+(\beta_2)^2$ is determined. That is all that can be determined
about $\beta_1$, $\beta_2$. In case $(iii)$, the Hamiltonian contains
$\beta_1$, $\beta_2$ only in $\beta_1 \pxi1 + \beta_2 \pxi2$. It is not
changed if $\beta_1$, $\beta_2$ are changed as the components of a
two-dimensional vector that is rotated when $\pxi1$, $\pxi2$ are changed
the same way. The dynamics is not changed for either qubit or for the
system of two qubits by this change of the $\pxi1$, $\pxi2$ used to describe the
$\Xi $ qubit. In case $(iii)$, there are $\pxi1$, $\pxi2$ for
which the dynamics is described by the Hamiltonian (\ref{eq:hamil}) with
$\beta_1$ positive and $\beta_2$ zero. Hence we assume that is the case. 

In sixth order, in $-i [ \; , \, H]$ applied to $\psig1$ six times, there
is a term $-(\gamma_3)^2 (\beta_1)^2 (\beta_3)^2 \psig1$ that, for case
$(iii)$, determines $(\beta_3)^2$ when $(\beta_1)^2$ is not zero. Case
$(iii)$ with $\beta_1$, $\beta_2$ both zero is the one exception overall.
It is described in Section \ref{sec5}.

\section{Signs \label{sec4}}

All the magnitudes of the parameters $\alpha$, $\beta$, $\gamma$ are
determined in all cases except the one described in Sec.~\ref{sec5}. The
signs of $\alpha_1$, $\alpha_2$, $\alpha_3$ also are determined. We can
change the signs of any two of $\pxi1$, $\pxi2$, $\pxi3$. That allows us to
choose $\pxi1$, $\pxi2$ $\pxi3$ so that neither $\gamma_2$ nor $\gamma_3$
is negative. Hence we assume that is the case. Then the sign of $\gamma_1$ is
the sign of $\vec \gamma_1 \cdot \vec \gamma_2 \times \vec \gamma_3 $. The sign
of $\beta_1$ is determined in cases $(i)$ and $(ii)$ because $\gamma_2 \gamma_3
\beta_1$ is determined. We are taking $\beta_1$ to be positive in case $(iii)$.
Only the signs of  $\gamma_1$, $\beta_2$, $\beta_3$ are not determined.

Can signs of $\gamma_1$, $\beta_2$, $\beta_3$ be changed without changing
the time dependence of states of the $\Sigma $ qubit? In case $(i)$, if the
sign of any one of $\gamma_1$, $\beta_2$, $\beta_3$ is changed, the signs
of all three must be changed, because $\gamma_1 \gamma_2 \beta_3$,
$\gamma_3 \gamma_1 \beta_2$ and $\gamma_2 \gamma_3 \beta_2 \beta_3$ are
determined. In case $(ii)$, if the sign of $\beta_2$ or $\beta_3$ is
changed, the signs of both must be changed, because $\gamma_2 \gamma_3
\beta_2 \beta_3$ is determined. In case $(iii)$, we are taking $\beta_2$ to
zero, so the sign of $\beta_3$ is the only one that can be changed. In all
cases, the only change that can be made is the change of signs of all the
$\gamma_1$, $\beta_2$, $\beta_3$ that are not zero.

Changing the signs of $\gamma_1$, $\beta_2$, $\beta_3$ does not change the
time dependence of states of the $\Sigma $ qubit. This change of signs
relates two different Hamiltonians that describe different dynamics for the
two qubits but give the same time dependence for states of the $\Sigma $
qubit. We will show this in two steps. In this section we show that the two
Hamiltonians give the same $u_{nj}(t)$, because the change of signs makes
no difference in any $\psig{j}$ terms in the power series for the $e^{itH}
\psig{n} e^{-itH}$. In Sec.~\ref{sec6} we will show that the two
Hamiltonians give the same $v_{nk}(t)\apxi{k}$ and
$w_{njk}(t)\apsigxi{j}{k}$.

Let $M$ be one of the $\psig{j}$, $\pxi{k}$, or $\psig{j} \pxi{k}$.
Consider one of the times that $M$ occurs in a power series for a $e^{itH}
\psig{n} e^{-itH}$. There are powers of the parameters $\alpha$, $\beta$,
$\gamma$ multiplying $M$. Let $p$ be the power of $\gamma_1$ plus the power
of $\beta_2$ plus the power of $\beta_3$. Now consider all the times that
$M$ occurs in the power series for the $e^{itH} \psig{n} e^{-itH}$. For
each $M$, either $p$ is even every time $M$ occurs or $p$ is odd every time
$M$ occurs. It is even for
\begin{equation}
\psig{j}, \quad \psig{j} \pxi{2}, \quad \psig{j} \pxi{3}, \quad \pxi1,
\label{eq:blue}
\end{equation}
which we call blue operators, and odd for
\begin{equation}
\psig{j} \pxi1, \quad \pxi2, \quad \pxi3,
\label{eq:red}
\end{equation}
which we call red operators. To see how this happens, consider how $p$ can
change. The power series are generated by repeated application of $[\; , \,
H]$. Each $[\; , \, \gamma_1 \psig1 \pxi1]$ brings in a power of
$\gamma_1$, each $[\; , \, \beta_2 \pxi2]$  a power of $\beta_2$, and each
$[\; , \, \beta_3 \pxi3]$ a power of $\beta_3$, so $p$ increases by $1$
with each $[\; , \, G]$ where $G$ is a term of $H$ that is a parameter
times a red operator, and $p$ does not change when $G$ is a parameter times
a blue operator. The commutator of two blue operators is a blue operator,
the commutator of two red operators is a blue operator, and the commutator
of a blue operator and a red operator is a red operator. The $[\; , \, G]$
that change $p$ are the $[\; ,\, G]$ that take blue operators to red
operators and red operators to blue operators, and the $[ \; , \, G]$ that
do not change $p$ are those that take blue to blue and red to red. A change
of $p$ between even and odd is a change of color. When an $M$ recurs, its
$p$ has changed between even and odd an even number of times. For each $M$,
either $p$ is even every time $M$ occurs or $p$ is odd every time $M$
occurs. Since $p$ is zero for the $\psig{n}$ at the start, $p$ must be even
for the blue operators and odd for the red operators. 

In particular, $p$ is even for the $\psig{j}$. The $\psig{j}$ terms are not
changed by the change of the signs of $\gamma_1$, $\beta_2$, $\beta_3$.
The $\psig{j}$ terms, the $u_{nj}(t)$, cannot distinguish the two
possibilities for the dynamics of the two qubits.

\section{The exception \label{sec5}}

The one exception is represented by case $(iii)$ with $\beta_1$, $\beta_2$
both zero. Then the Hamiltonian is
\begin{equation}
H = \alpha_j \psig{j} + \gamma_3 \psig3 \pxi3 + \beta_3 \pxi3.
\label{eq:hamilreduced}
\end{equation}
Here $\beta_3$ can vary over the whole range of real numbers without
changing the time dependence of states of the $\Sigma $ qubit; the $\beta_3
\pxi3$ term commutes with the rest of $H$ and with $\psig1$, $\psig2$,
$\psig3$.

\section{Two-qubit states \label{sec6}}

The mean values $\apxi{k}$ and $\apsigxi{j}{k}$ describe the initial state
of the $\Xi $ qubit and of the correlations between the two qubits. They
also are generally determined to within a change of signs by the time
dependence of states of the $\Sigma $ qubit. We have two possibilities for
the dynamics of the two qubits. For each possibility, the parameters
$\alpha$, $\beta$, $\gamma$ are determined by the time dependence of states
of the $\Sigma $ qubit, so the Hamiltonian $H$ and the $v_{nk}(t)$,
$w_{njk}(t)$ are determined. The mean values $\apsig{n}(t)$ that describe
states of the $\Sigma $ qubit in time provide linear equations
(\ref{eq:mean}) for the $\apxi{k}$, $\apsigxi{j}{k}$. The twelve
$\apxi{k}$, $\apsigxi{j}{k}$ generally are determined by the four time
derivatives in the power series for the $\big\langle e^{itH} \psig{n}
e^{-itH} \big\rangle$ to fourth order for the three values of $n$. There
are many exceptions. For example, in the case we have described
\cite{jordan04a} where $\gamma_3$ is the only one of the $\alpha$, $\beta$,
$\gamma$ that is not zero, only $\apsigxi{1}{3}$ and $\apsigxi{2}{3}$ are
determined, and when only $\gamma_2$ and $\gamma_3$ are nonzero, only
$\apxi{1}$, $\apsigxi{1}{2}$, $\apsigxi{1}{3}$, $\apsigxi{2}{3}$,
$\apsigxi{3}{2}$ are determined; the other $\apxi{k}$ and $\apsigxi{j}{k}$
have no effect on the time dependence of states of the $\Sigma $ qubit. 

The $\apxi{k}$, $\apsigxi{j}{k}$ for one possibility for the dynamics of
the two qubits are changed to the $\apxi{k}$, $\apsigxi{j}{k}$ for the
other possibility by just changing the signs of the $\apsigxi{j}{1}$,
$\apxi{2}$, $\apxi{3}$. This follows from what we learned in
Sec.~\ref{sec4}. Let $M$ be one of the $\pxi{k}$ or $\psig{j} \pxi{k}$.
Every time $\langle M \rangle$ occurs in the power series for the equations
(\ref{eq:mean}) it is multiplied by powers of the $\alpha$, $\beta$,
$\gamma$ for which $p$ is even if $M$ is a blue operator, odd if $M$ is a
red operator. Changing from one possibility for the dynamics to the other,
changing the signs of $\gamma_1$, $\beta_2$, $\beta_3$, changes the
equations for the $\apxi{k}$, $\apsigxi{j}{k}$ by just changing the signs
of the terms for which $p$ is odd. That just changes the signs of the
coefficients of the $\langle M \rangle$ for which $M$ is a red operator,
the coefficients of the $\apsigxi{j}{1}$, $\apxi{2}$, $\apxi{3}$. The
equations for the $\apxi{k}$, $\apsigxi{j}{k}$ for one possibility for the
dynamics are the same as the equations for the other possibility for
$\apxi{k}$, $\apsigxi{j}{k}$ with the signs of the $\apsigxi{j}{1}$,
$\apxi{2}$, $\apxi{3}$ changed.

Now we can see that the time dependence of states of the $\Sigma $ qubit
does not distinguish the two possibilities for the dynamics of the two
qubits. The mean values $\apsig{n}(t)$ that describe the states of the
$\Sigma $ qubit in time provide Eqs.~(\ref{eq:mean}). Some of the terms of
the power series for these equations generally determine the $\apxi{k}$ and
$\apsigxi{j}{k}$. When these $\apxi{k}$ and $\apsigxi{j}{k}$ are used in
the remaining terms, the equations obtained are the same for the two
possibilities for the dynamics of the two qubits. Again let $M$ be one of
the $\pxi{k}$ or $\psig{j} \pxi{k}$. If $M$ is a blue operator, then both
$\langle M \rangle$ and the powers of $\alpha$, $\beta$, $\gamma$ that
multiply it in these equations are the same for the two possibilities. If
$M$ is a red operator, then both $\langle M \rangle$ and the powers of
$\alpha$, $\beta$, $\gamma$ that multiply it have the same magnitude and
opposite signs for the two possibilities. The ability of
Eqs.~(\ref{eq:mean}) to distinguish the two possibilities is not increased
in the exceptional cases where some of the $\apxi{k}$ or $\apsigxi{j}{k}$
are not determined.

\section{Example \label{sec67}}

Here is a substantial example. Suppose $\alpha_1$, $\alpha_2$, $\alpha_3$ and
$\beta_1$, $\beta_2$, $\beta_3$ are zero. Then the dynamics generated by the
Hamiltonian (\ref{eq:hamil}) can be worked out very simply \cite{jordan06a}. The
three matrices $\psig 1 \pxi 1$, $\psig 2 \pxi 2$, $\psig 3 \pxi 3$ commute with
each other (The different $\psig j$ anticommute and the different $\pxi j$
anticommute, so the different $\psig j \pxi j$ commute). That allows us to
easily compute
\begin{eqnarray}
\label{eq:int2}
\langle e^{itH}\psig 1 e^{-itH} \rangle & = & \langle \psig 1 e^{-i \gamma_2
t\psig 2 \pxi 2} e^{-i \gamma_3 t\psig 3 \pxi 3} \rangle \nonumber \\
&=& \langle \psig 1 \rangle \cos \gamma_2 t\cos \gamma_3 t  \nonumber \\
&& \hspace{2 mm} + \langle \pxi 1 \rangle \sin \gamma_2 t\sin \gamma_3 t
\nonumber \\
&& \hspace{4 mm} - \langle \psig 2 \pxi 3 \rangle \cos \gamma_2 t \sin \gamma_3
t \nonumber \\
&& \hspace{6 mm} + \langle \psig 3 \pxi 2 \rangle \sin \gamma_2 t \cos \gamma_3
t \qquad
\end{eqnarray}
using the algebra of Pauli matrices, and similarly
\begin{eqnarray}
\label{eq:int3}
\langle e^{itH}\psig 2 e^{-itH} \rangle &=& \langle \psig 2 \rangle \cos
\gamma_3 t \cos \gamma_1 t \nonumber \\
&& \hspace{2 mm} + \langle \pxi 2 \rangle \sin \gamma_3 t \sin
\gamma_1 t \nonumber \\
&& \hspace{4 mm} - \langle \psig 3 \pxi 1 \rangle \cos \gamma_3 t \sin \gamma_1
t \nonumber \\
&&  \hspace{6 mm}+ \langle \psig 1 \pxi 3 \rangle \sin \gamma_3 t\cos \gamma_1
t, \qquad
\end{eqnarray}
\begin{eqnarray}
\label{eq:int4}
\langle e^{itH}\psig 3 e^{-itH} \rangle &=& \langle \psig 3 \rangle \cos
\gamma_1 t \cos \gamma_2 t \nonumber \\
&&  \hspace{2 mm}+ \langle \pxi 3 \rangle \sin \gamma_1 t\sin \gamma_2
t \nonumber \\
 && \hspace{4 mm} - \langle \psig 1 \pxi 2 \rangle \cos \gamma_1 t\sin \gamma_2
t  \nonumber \\
&&  \hspace{6 mm}+ \langle \psig 2 \pxi 1 \rangle \sin \gamma_1 t\cos \gamma_2
t. \qquad
\end{eqnarray}
These Eqs.~(\ref{eq:int2}) - (\ref{eq:int4}) give examples of the functions
$u_{nj}(t)$, $v_{nk}(t)$, $w_{njk}(t)$ in Eqs.(\ref{eq:mean}). We consider the
cases where $\gamma_1$, $\gamma_2$, $\gamma_3$ are all nonzero.

The way time dependence for the one qubit reveals the dynamics of the two is
particularly clear in these cases. The $u_{nj}(t)$ are even functions of $t$, so
their power series have no first-order or third-order terms. This is not changed
when the $\Sigma_j $ and the $\Xi_k $ are changed by rotations. Therefore, for
any $\Sigma_j $ and $\Xi_k $ that are used at the start, the results of the
procedures described in the first and sixth paragraphs of Section \ref{sec3}
are that $\alpha_1$, $\alpha_2$, $\alpha_3$ are zero and that $\beta_1$,
$\beta_2$, $\beta_3$ are zero in the cases where $\gamma_1$, $\gamma_2$,
$\gamma_3$ are all nonzero. In between, the procedure described in the second
paragraph of Section \ref{sec3} determines the dot products (\ref{dots}), so the
Hamiltonian can be put in the form (\ref{eq:hamil}). The magnitudes
$(\gamma_1)^2$, $(\gamma_2)^2$,  $(\gamma_3)^2$ are determined. The signs of
$\gamma_1$, $\gamma_2$, $\gamma_3$ are not determined. The Pauli matrices
$\pxi1$, $\pxi2$, $\pxi3$ can be chosen so  that $\gamma_2$ and $\gamma_3$ are
positive. This leaves the sign of $\gamma_1$ undetermined. It is the sign of
$\vec \gamma_1 \cdot \vec \gamma_2 \times \vec \gamma_3 $.

Thus, we see that time dependence for the one qubit reveals that the
Hamiltonian can be put in the form (\ref{eq:hamil}) and that $\alpha_1$,
$\alpha_2$, $\alpha_3$ and $\beta_1$, $\beta_2$, $\beta_3$ are zero for the
cases where $\gamma_1$, $\gamma_2$, $\gamma_3$ are all nonzero. This gives
Eqs.(\ref{eq:int2}) - (\ref{eq:int4}). There are two different possibilities for
the dynamics of the two qubits, corresponding to the two different signs of
$\gamma_1$. We can also see that for each sign of $\gamma_1$, the
Eqs.(\ref{eq:int2}) - (\ref{eq:int4}) give the $\apxi{k}$ and the
$\apsigxi{j}{k}$ for $j\not= k$. The results for the two different signs of
$\gamma_1$ differ in the signs of $\apsigxi{2}{1}$, $\apsigxi{3}{1}$ and
$\apxi{2}$, $\apxi{3}$, as described in Section \ref{sec6}.

The three mean values $\apsigxi{k}{k}$ are not determined. They are constants
of the motion and they are not involved with the time dependence of the other
mean values for the two qubits.

In this example it is also particularly clear that there is a real difference
between the two possibilities for the dynamics of the two qubits corresponding
to the two different signs of $\gamma_1$. There is no unitary transformation
that changes one into the other. Changing the sign of $\gamma_1$ changes the
eigenvalue spectrum of the Hamiltonian. The Hamiltonian (\ref{eq:hamil}) for
these cases is a function
\begin{equation}
H = \frac{1}{2}[-\gamma_1 (\psig{2}\pxi{2})(\psig{3}\pxi{3}) + \gamma_2
(\psig{2}\pxi{2}) + \gamma_3 (\psig{3}\pxi{3})]
 \label{eq:hamilex}
\end{equation}
of the two matrices $\psig{2}\pxi{2}$ and $\psig{3}\pxi{3}$ which form a
compete set of commuting operators. The four pairs of their eigenvalues label a
basis of eigenvectors $|1, 1\rangle $, $|1, -1\rangle $, $|-1, 1\rangle $, $|-1,
-1\rangle $ for the space of states of the two qubits. This shows that the
eigenvalues of $H$ are
\begin{eqnarray}
&\frac{1}{2}[-\gamma_1 + \gamma_2 + \gamma_3 ] \nonumber \\
&\frac{1}{2}[\gamma_1 + \gamma_2 - \gamma_3 ] \nonumber \\
&\frac{1}{2}[\gamma_1 - \gamma_2 + \gamma_3 ] \nonumber \\
&\frac{1}{2}[-\gamma_1 - \gamma_2 - \gamma_3 ].
\end{eqnarray}

For positive $\gamma_1$, there is an eigenvalue of $H$ that is less than any
eigenvalue of $H$ for negative $\gamma_1$. For negative $\gamma_1$, there is an
eigenvalue of $H$ that is greater than any eigenvalue of $H$ for positive
$\gamma_1$.

\section{Classical analog \label{sec7}}

To see how much quantum mechanics is involved, we consider a classical
analog. It exhibits the logical structure of our method in a setting where it
can be easily seen. Let $x$ and $y'$ be real variables that have linear
equations of motion
\begin{equation}
\frac{d x}{dt} = \alpha x + \gamma' y' , \quad \frac{dy'}{dt} = \delta x +
\beta y'
\label{eq:class1}
\end{equation}
with real parameters $\alpha$, $\beta$, $\gamma'$, $\delta$. We assume
that neither $\gamma'$ nor $\delta$ is zero, and if $\gamma'$ is negative
we change the signs of $\gamma'$, $\delta$ and $y'$ to make $\gamma'$
positive. Let
\begin{equation}
y =\sqrt{\frac{\gamma'}{|\delta|}}y' , \quad \gamma =\sqrt{\gamma' |\delta
|}.
\label{eq:class2}
\end{equation}
Then
\begin{equation}
\frac{dx}{dt} = \alpha x + \gamma y , \quad \frac{dy}{dt} = \pm \gamma x +
\beta y
\label{eq:class3}
\end{equation}
with the $\pm$ the sign of $\delta$, and
\begin{eqnarray}
\frac{d^2 x}{dt^2} &=& \alpha^2 x \pm \gamma^2 x + \gamma(\alpha + \beta) y
\nonumber \\
\frac{d^3 x}{dt^3} &=& (\alpha^3 \pm 2 \alpha \gamma^2 )x \pm \gamma^2
\beta x  \nonumber \\
&& \qquad + (\alpha^2 \gamma + \alpha \beta \gamma + \beta^2 \gamma
\pm \gamma^3 )y .
\label{eq:class4}
\end{eqnarray}
Looking at the time derivatives of $x$ for variable $x$, we learn $\alpha$
from $dx/dt$, then $\gamma^2$ and the $\pm$ sign from $dx^2/dt^2$, and
$\beta$ from $d^3x/dt^3$. The parameters in the equations of motion for $x$
and $y$ are determined from the power series for the time dependence of $x$
for variable initial values of $x$. The initial value of $y$ is determined
from the terms in the derivatives of $x$ that do not depend on $x$. This is
closely analogous to the calculations of Sec.~\ref{sec3}.

When neither $\gamma'$ nor $\delta$ is zero, there is a $y$ for which the
dynamics of $x$ and $y$ is described by the equations of motion
(\ref{eq:class3}). Then there is feedback from the equation of motion for
$y$ to the time dependence of $x$ that depends on the initial value of $x$
and allows all the parameters in the equations of motion for $x$ and $y$ to
be determined from the time dependence of $x$ alone.

For qubits, we needed quantum mechanics just to identify qubits as the
simplest objects to consider and to specify the form that equations of
motion for two qubits can take. Everything we learned can be obtained from
the equations of motion for the mean values $\apsig{j}$, $\apxi{k}$,
$\apsigxi{j}{k}$. There is no reason in principle that these could not be
classical equations of motion for some system. They could even be
Hamiltonian equations, with a Hamiltonian function of the form
(\ref{eq:hamil}), for classical spin variables with suitably defined
Poisson brackets \cite{ecgs74a}.

\section{Discussion \label{sec8}}

We have shown that the Hamiltonian for two qubits can be determined almost
completely from the time dependence of states of one of the qubits, but it
requires rather detailed knowledge of that time dependence. In our
calculations we generally need time derivatives up to third order, and in
some cases up to sixth order. A practical application would be difficult.

The result seems to raise a question; it suggests that a step might remain
to be taken to understand the mathematical result physically. Why is the
time dependence of states of the one qubit unchanged when the Hamiltonian
for the two qubits is changed in just those signs of three terms? Is there
a reason, a physical explanation, perhaps a symmetry, that would let us
predict this result without doing the calculations? Is there more to be
said about that change of three signs? We have not found an answer. It may
be that this result is just what happens mathematically in this particular
situation, not a consequence of something more general, more physical, or
more easily understood.

\appendix*

\section{Diagonalizing the Gammas}

Here is a simple proof that rotations can diagonalize the matrix of
coefficients $\gamma_{jk}$ for the interaction terms and put the Hamiltonian in
the form (\ref{eq:hamil}), and that the rotation of the $\psig{j}$ and $u_{nj}(t)$ that can be found
with knowledge of the dot products (\ref{dots}). The $\gamma $ vectors are central. Their dot products
$\vec \gamma_m \cdot \vec \gamma_n $ are the elements of a real symmetric
$3\times 3$ matrix, so there is a real $3\times 3$ rotation matrix $R$ with
elements $R_{jm}$ such that
\begin{equation}
\label{orthogonal}
R_{jm} \vec \gamma_m \cdot R_{kn} \vec \gamma_n = R_{jm} \,\vec \gamma_m \cdot
\vec \gamma_n \, (R^{-1})_{nk} = 0
\end{equation}
for $j\not = k$. The three vectors $R_{jm} \vec \gamma_m $ are orthogonal.
There is a rotation $S$ that changes these three vectors and the $\pxi1$,
$\pxi2$, $\pxi3$ together, as described in Section \ref{sec3}, and takes each
$R_{jm} \vec \gamma_m $ to a vector $SR_{jm} \vec \gamma_m $ that is along the
$j$ axis. Then
\begin{eqnarray}
\label{diagonal}
\gamma_{jk} \psig{j}\pxi{k} & = & \psig{j}\vec \gamma_j \cdot \vec \Xi
\nonumber \\
 & = & (R_{jm} \Sigma_m ) R_{jn} \vec \gamma_n \cdot \vec \Xi \nonumber \\
 & = & (R_{jm} \Sigma_m ) SR_{jn} \vec \gamma_n \cdot S\vec \Xi\nonumber \\
 & = & (R_{jm} \Sigma_m ) \gamma_j (S\vec \Xi )_j \nonumber \\
 & = & \gamma_j(R_{jm} \Sigma_m )(S_{jl} \Xi_l )
\end{eqnarray}
where $\gamma_j $ is the only component of $SR_{jm} \vec \gamma_m $ that may be
nonzero, the component in the $j$ direction, which may be positive, negative, or
zero, and the $S_{jk} $ are the elements of the rotation matrix for $S$.

The $\psig1$, $\psig2$, $\psig3$ are rotated by $R$ and the $\pxi1$, $\pxi2$,
$\pxi3$ are rotated by $S$. The rotation $R$ depends only on the dot products
$\vec \gamma_m \cdot \vec \gamma_n $, which are determined by the time dependence of states of
the $\Sigma $ qubit. The $\gamma $ vectors are changed differently by the
two rotations. The vector $SR_{jm} \vec \gamma_m $ is just the vector $R_{jm}
\vec \gamma_m $ rotated by $S$. The vector $R_{jm} \vec \gamma_m $ is not the
vector $\vec \gamma_m $ rotated by $R$; it is the linear combination with
coefficients $R_{j1} $, $R_{j2} $, $R_{j3} $ of the vectors $\vec \gamma_1 $,
$\vec \gamma_2 $, $\vec \gamma_3 $.

There is no change in the value of $\vec \gamma_1 \cdot \vec \gamma_2 \times
\vec \gamma_3 $, only a change of what it is called. It is
\begin{eqnarray}
\label{gamma123}
\gamma_1 \gamma_2 \gamma_3 & = & (SR_{1l} \vec \gamma_l )\cdot (SR_{2m} \vec
\gamma_m )\times (  SR_{3n} \vec{\gamma}_n ) \nonumber \\
 & = & (R_{1l} \vec \gamma_l )\cdot (R_{2m} \vec \gamma_m )\times 
(R_{3n}\vec \gamma_n ) \nonumber \\
 & = & (R_{11}R_{22}R_{33}-R_{11}R_{23}R_{32} \nonumber \\ 
&& \hspace{1 mm}  +R_{12}R_{23}R_{31} - R_{12}R_{21}R_{33} \nonumber \\
&& \hspace{2 mm} +R_{13}R_{21}R_{32}-R_{13}R_{22}R_{31})\vec \gamma_1
\cdot \vec \gamma_2 \times \vec \gamma_3\nonumber \\
 & = & \det(R)\, \vec \gamma_1 \cdot \vec \gamma_2 \times \vec \gamma_3
\nonumber \\
 & = & \vec \gamma_1 \cdot \vec \gamma_2 \times \vec \gamma_3
\end{eqnarray}
because $\det(R)$ is $1$ for a rotation.

\acknowledgments
We are grateful to a referee for very helpful questions and suggestions,
including the idea of the example. Anil Shaji Acknowledges the support of US
Office of Naval Research Contract
No. N00014-03-1-0426

\bibliography{/apu/shaji/Documents/tex/bib/anilbib}

\end{document}